\documentclass[twocolumn,secnumarabic,amssymb, nobibnotes, aps, prd, superscriptaddress]{revtex4}

\usepackage[utf8]{inputenc}
\usepackage{graphics,graphicx}
\usepackage{amsmath,amssymb,amsfonts,latexsym,cancel}
\usepackage{bm}

\begin{document}

\title{Extra dimensions' influence on the equilibrium and radial stability of  strange quark stars}

\author{Jos\'e D. V. Arba\~nil}
\email{jose.arbanil@upn.pe}
\affiliation{Departamento de Ciencias, Universidad Privada del Norte, Avenida el Sol 461 San Juan de Lurigancho, 15434 Lima,  Peru}
\author{Geanderson A. Carvalho}
\email{araujogc@ita.br}
\affiliation{Departamento de F\'isica, Instituto Tecnol\'ogico de Aeron\'autica, Centro T\'ecnico Aeroespacial, 12228-900 S\~ao Jos\'e dos Campos, S\~ao Paulo, Brazil}
\affiliation{Dipartimento di Fisica, Sapienza Universit\`a di Roma, Piazzale Aldo Moro 5, I-00185 Rome, Italy}
\affiliation{ICRANet, Piazza della Repubblica 10, I-65122 Pescara, Italy}
\author{Ronaldo V. Lobato}
\email{rvlobato@ita.br}
\affiliation{Departamento de F\'isica, Instituto Tecnol\'ogico de Aeron\'autica, Centro T\'ecnico Aeroespacial, 12228-900 S\~ao Jos\'e dos Campos, S\~ao Paulo, Brazil}
\affiliation{Dipartimento di Fisica, Sapienza Universit\`a di Roma, Piazzale Aldo Moro 5, I-00185 Rome, Italy}
\affiliation{ICRANet, Piazza della Repubblica 10, I-65122 Pescara, Italy}
\author{Rubens M. Marinho Jr.}
\email{marinho@ita.br}
\affiliation{Departamento de F\'isica, Instituto Tecnol\'ogico de Aeron\'autica, Centro T\'ecnico Aeroespacial, 12228-900 S\~ao Jos\'e dos Campos, S\~ao Paulo, Brazil}
\author{Manuel Malheiro}
\email{malheiro@ita.br}
\affiliation{Departamento de F\'isica, Instituto Tecnol\'ogico de Aeron\'autica, Centro T\'ecnico Aeroespacial, 12228-900 S\~ao Jos\'e dos Campos, S\~ao Paulo, Brazil}

\date{\today}

\begin{abstract}
We analyze the influence of extra dimensions on the static equilibrium configurations and stability against radial perturbations. For this purpose, we solve  stellar structure equations and radial perturbation equations, both modified for a $d$-dimensional spacetime ($d\geq4$) considering that  spacetime outside the object is described by a Schwarzschild-Tangherlini metric. These equations are integrated considering a MIT bag model equation of state extended for $d\geq4$. We show that the spacetime dimension influences both the structure and stability of compact objects.  For an interval of central energy densities $\rho_{cd}\,G_d$ and total masses $MG_d/(d-3)$, we show that the stars gain more stability when the dimension is increased. In addition, the maximum value of $M{G_d}/(d-3)$ and the zero eigenfrequency of oscillation are found with the same  value of $\rho_{cd}\,G_d$; i.e., the peak value of $M{G_d}/(d-3)$ marks the onset of instability. This indicates that the necessary and sufficient conditions to recognize regions constructed by stable and unstable equilibrium configurations against radial perturbations are, respectively, $dM/d\rho_{cd}>0$ and $dM/d\rho_{cd}<0$. We obtain that some physical parameter of the compact object in a $d$-dimensional spacetime, such as the radius and the mass, depend of the normalization. Finally, within the Newtonian framework, the results show that compact objects with adiabatic index  $\Gamma_1\geq2(d-2)/(d-1)$ are stable against small radial perturbations. 
\end{abstract}
\maketitle


\section{Introduction}\label{sec-introd}

In recent decades, as a direct consequence of  Kaluza-Klein theory \cite{kaluza1921,klein1926} and some other theory on supergravity, the idea that spacetime may have extra dimensions, as yet undetected by experiment, has become accepted. Motivated by this idea, the implications of the extra dimensions on some physical phenomena arising in the study of compact objects is theoretically investigated. For instance, within the frame of the Newtonian theory of gravity, some studies have analyzed the static equilibrium configurations of white dwarfs \cite{chavanis2007,bechhoefer_chabrier1993}. Moreover, within the general relativity (GR) context, the influence of the spacetime dimension in the equilibrium configuration of compact objects \cite{harko_mak2000,poncedeleon2000}, the gravitational collapse of compact stars \cite{ghosh_beesham2001,ghosh_dadhich2001,ghosh_beesham2000}, and the properties black holes \cite{myers_perry1986,emparan_reall2008,gibbons_ida_shiromizu} have been addressed.

In Newtonian gravity, the most complete investigations on static equilibrium configurations of compact stars, in a higher-dimensional spacetime, have been investigated in Refs.~ \cite{chavanis2007,bechhoefer_chabrier1993}. In these articles, based on the seminal paper of Chandrasekhar \cite{chandra3}, it has been shown that white dwarfs become unstable in higher dimensions due to the pressure of degenerate fermions that cannot counterbalance the gravitational action, starting a gravitational collapse or evaporating. Consequently, the fermion stars \cite{chavanis2007,bechhoefer_chabrier1993} cannot exist in spacetime dimensions larger than $d=4$. This implies that extra dimensions, if they exist, need to be compactified since white dwarfs are confirmed by astronomical observations. In addition, these investigations considered that, at a star's surface, the energy density vanishes together with the pressure. In the case of quark stars, fermion stars in which the pressure goes to zero at the surface but the energy density does not, one naturally questions the validity of such conclusions. Thus, in the frame of GR, we investigate the static structure configuration and the radial stability of a compact object in a higher-dimensional spacetime by taking into account a MIT bag model equation of state (EOS) in $d$-dimensions.

In GR, the influence of the dimension in the static equilibrium configurations of homogeneous compact objects  \cite{harko_mak2000,poncedeleon2000} is investigated. This is realized by using the Tolman-Oppenheimer-Volkoff (TOV) equation, also known as the hydrostatic equilibrium equation, for a higher-dimensional spacetime. In Refs.~\cite{harko_mak2000,poncedeleon2000}, by using different normalization factors, it is shown how the total mass depends of the spacetime dimension. In the latter work, Ref. \cite{poncedeleon2000}, using the mass (energy) conservation equation for a higher-dimensional spacetime, the authors explain that the dimensionality of spacetime increases the mass of the object; however, it does not contribute to the fluid pressure. Consequently, in a fixed volume where the fluid pressure is fixed, a compact object would be more prone to gravitational collapse in dimensions greater than $4$. 

It is well known that gravitational collapse occurs when an object's internal pressure cannot sustains its own gravity, yielding to formations of new stellar structures. Depending of the initial conditions imposed, the final result of such gravitational collapse could be either a naked singularity or a black hole.

The gravitational collapse has been investigated in a wide range of scenarios, including higher-dimensional spacetime. Among these, it is highlighted how the extra dimension affects the gravitational collapse of an inhomogeneous dust cloud \cite{ghosh_beesham2001,ghosh_beesham2000} and null fluid \cite{ghosh_dadhich2001}. In fact, such contributions suggest that the naked singularity diminishes with the increase of the spacetime dimension, which is a signature of the collapse of an object with infinity density. Thus, it would be less likely as this spacetime dimension increases.  

In contrast, in a black hole, where the singularity is completely surrounded by an event horizon, unlike to naked singularity, the singularity cannot be  observed. Moreover, Refs.~\cite{myers_perry1986,emparan_reall2008,gibbons_ida_shiromizu} analyze the implication of spacetime dimensions such as the singularity, topology, and dynamical stability, among other factors.

We investigate the influence of the extra dimensions in the static structure and radial stability of compact objects considering a MIT bag model EOS in $d$ dimensions in the framework of GR. This study is analyzed by solving numerically the hydrostatic equilibrium \cite{tolman,oppievolkoff} and the Chandrasekhar radial pulsation equation \cite{chandrasekhar_rp,chandrasekhar_PRL}, fully including the extra dimension effects.  We also analyze the dependence of some  physical compact object properties such the fluid pressure, mass, radius, compressibility factor, redshift, and the fundamental mode eigenfrequency with the spacetime dimensions. Moreover, we discuss the change of such parameters with the normalization considered in this work and those assumed in Refs.~\cite{harko_mak2000,poncedeleon2000}. Finally, we study the stability of these compact objects in the Newtonian frame.

This paper is divided as follows: In Sec.~\ref{sec-basicequations}, the general relativistic formulations are presented; the steps to follow to derive both the stellar structure and the stellar stability equations and equation of state are shown. Section \ref{results} shows how the hydrostatic equilibrium and radial stability are affected by both the extra dimensions and the normalization chosen. In this section, we also study the radial stability of these objects in the framework of Newtonian gravity. Finally, in Sec.~\ref{conclusion}, we conclude. 

Throughout this article, we consider the speed of light $c$ and the four-dimensional gravitational constant $G_4$ to be equal to unity; i.e., $c=1=G_4$.

\section{General relativistic formulation in $d$ dimensions}\label{sec-basicequations}

\subsection{Einstein field equation}

The properties of compact objects in higher dimensions are analyzed as a description of the Einstein equation in $d$ dimensions, $d\geq4$. For that purpose, the field equation in $d$ spacetime dimensions is assumed to be of the form \cite{lemosezanchinbonnor}
\begin{equation}\label{einstein_tensor}
G_{\mu\nu} = \frac{d-2}{d-3}S_{d-2}G_d T_{\mu\nu}\, ,
\end{equation}
with $\displaystyle{G_{\mu\nu}=R_{\mu\nu}-\frac{1}{2}g_{\mu\nu} R}$ being the Einstein tensor and the quantities $R_{\mu\nu}$, $R$ and $g_{\mu\nu}$ representing, respectively, the Ricci tensor, the Ricci scalar and the metric tensor. The right-hand side of Eq.~(\ref{einstein_tensor}) bears the universal constant $G_d$, which in four dimensions corresponds to the Newton's gravitational constant. $S_{d-2}=2\pi^{(d-1)/2}/\Gamma((d-1)/2)$ is the area of unitary sphere $\mathbf{S}^{{d}-2}$, where $\Gamma$ is the usual gamma function, and the factor $(d-2)\,G_d S_{d-2}/(d-3)$ corresponds to the $8\pi$ term in four dimensions (see Ref.~\cite{lemosezanchinbonnor}). $T_{\mu\nu}$ represents the matter energy-momentum tensor of a perfect fluid, which in this study is written as 
\begin{equation}\label{fluidemt}
T_{\mu\nu}= \left(\rho_{0d}+p_{0d}\right)U_\mu U_\nu +p_{0d}\,g_{\mu\nu},
\end{equation}
with $\rho_{0d}$ being the energy density, $p_{0d}$ the pressure of the fluid and $U_\mu$ the velocity of the fluid in the $d$-dimensional spacetime. Also, the fluid velocity follows the condition
\begin{equation}\label{velocity_d}
U_{\mu}U^{\mu}=-1.
\end{equation}
In all of the above definitions, the Greek indices $\mu,\nu$, etc., run from $0$ to $d-1$, where $0$ represents the time, and the other $d-1$ coordinates are spacelike.

\subsection{The background spacetime}

To describe the static hyperspherically symmetric distribution of the fluid, we consider the interior line element to be given by 
\begin{equation}\label{int_match}
ds^{2}=-e^{\nu_0}\,dt^2+e^{\lambda_0}\,dr^2+r^2\sum^{d-2}_{i=1}
\left(\prod^{i-1}_{j=1}\sin^{2}\theta_{j}\right)d\theta^{2}_{i},
\end{equation}
where the functions $\nu_0=\nu_0(t,r)$ and $\lambda_0=\lambda_0(t,r)$ depend on the temporal $t$ and the radial coordinate $r$.

It is worth mentioning that the functions considered in the metric potentials ($\nu_0$ and $\lambda_0$) and in the variables of the fluid ($p_{0d}$ and $\rho_{0d}$) depend on the temporal $t$ and the radial coordinate $r$. To analyze the radial perturbation in an equilibrium configuration, it is necessary to perturb the variables of the metric and fluid. Thus, following the method applied by Chandrasekhar in Ref.~\cite{chandrasekhar_rp}, we decompose the aforementioned variables into the form
\begin{equation}
f_0(t,r)=f(r)+\delta f(t,r),\label{relation}
\end{equation}
with $f(r)$ being the quantities that depend on the variable $r$, only. In turn, $\delta f(t,r)$ depicts the Eulerian perturbations that depend on the variables $t$ and $r$.

\subsection{The stellar structure equations}

The static equilibrium configurations are analyzed through the stellar structure equations. With the aim of deriving this set of equations, we consider relation \eqref{relation} and $\delta f(t,r)=0$ in the nonzero components of the Einstein field equation and the Bianchi identity, yielding
\begin{eqnarray}
\hspace{-0.4cm}&&\frac{1}{2e^{\lambda}r}\frac{d\lambda}{dr}-\frac{(d-3)}{2e^{\lambda}r^2}+\frac{(d-3)}{2r^2}=\frac{1}{d-3}S_{d-2}G_d\,\rho_{d},\label{G00_r}\\
\hspace{-0.4cm}&&\frac{1}{2e^{\lambda}r}\frac{d\nu}{dr}+\frac{(d-3)}{2e^{\lambda}r^2}-\frac{(d-3)}{2r^2}=\frac{1}{d-3}S_{d-2}G_d\, p_{d},\label{G11_r}\\
\hspace{-0.4cm}&&\frac{d\nu}{dr}=-\frac{2}{(p_{d}+\rho_{d})}\frac{dp_{d}}{dr}.\label{df/f}
\end{eqnarray}

Now, to represent the gravitational mass in a $d$-dimensional spacetime, the function $m\,G_d/(d-3)$ is introduced in such a way that:
\begin{equation}\label{g11}
e^{-\lambda}=1-\frac{2m\,G_{d}}{(d-3)r^{d-3}}.
\end{equation}
Replacing the last equality in Eq.~\eqref{G00_r}, we obtain:
\begin{equation}\label{dm/dr}
\frac{dm}{dr}=S_{d-2}\rho_{d} r^{d-2}\,.
\end{equation}
Equation \eqref{dm/dr} represents mass (energy) conservation as measured in the hypersphere frame.

Combining Eqs.~\eqref{G11_r} and \eqref{df/f}, the $d$-dimensional hydrostatic equilibrium equation is represented by:
\begin{equation}\label{tov}
\frac{dp_{d}}{dr}=-(p_{d}+\rho_{d})G_{d}\left[\frac{\frac{S_{d-2}p_{d}r}{(d-3)}+\frac{m}{r^{d-2}}}{1-\frac{2m\,G_{d}}{(d-3)r^{d-3}}}\right].
\end{equation}
Equation \eqref{tov} is also called as the Tolman-Oppenheimer-Volkoff equation. This equation is modified from its original form to include the influence of the spacetime dimensions \cite{poncedeleon2000}. In addition, for $d=4$, Eq.~(\ref{tov}) is reduced to the traditional TOV equation \cite{tolman,oppievolkoff}. 

In order to find stellar equilibrium configurations, the stellar structure equations, Eqs.~\eqref{df/f}, \eqref{dm/dr}, and \eqref{tov}, must be integrated, from the center toward the surface of the star.

The stellar structure equations integration starts in the center of the hypersphere ($r=0$), where
\begin{equation}
\begin{array}{l}
m(0)=0,\hspace{0.3cm}\lambda(0)=0,\hspace{0.3cm}\nu(0)=\nu_c,\\ 
p_{d}(0)\,G_d=p_{cd}\,G_d,\hspace{0.3cm}{\rm and}\hspace{0.3cm} \rho_{d}(0)\,G_d=\rho_{cd}\,G_d,\label{int_cond}
\end{array}
\end{equation}
up until the surface of the object ($r=R$), which is attained when the fluid pressure vanishes, i.e.,
\begin{equation}\label{pressure_r}
p_{d}(r=R)\,G_d=0.
\end{equation}
The variables $p_{c}\,G_d$ and $\rho_{c}\,G_d$ represent respectively, the pressure and energy density at the center of the object. 

At the object's surface, the interior spacetime connects smoothly to the exterior Schwarzschild-Tangherlini spacetime \cite{lemosezanchin2009,tangherlini1963}, where the interior and the exterior metric functions satisfy the relation:
\begin{equation}\label{metric_functions}
e^{\nu(R)}=\frac{1}{e^{\lambda(R)}}=1-\frac{2M\,G_{d}}{(d-3)r^{d-3}},
\end{equation}
with $MG_d/(d-3)$ being the total mass.

\subsection{The radial stability equations}

To investigate the spacetime dimension influences in the radial stability, Eulerian perturbations must be determined. As a first step, the $d$-dimensional velocity components are defined. With aim to satisfying Eq.~\eqref{velocity_d}, these components have the same form as those used by Chandrasekhar \cite{chandrasekhar_rp}.  Later, both the metric functions and fluid properties are decomposed into the form indicated in Eq.~\eqref{relation}. The definitions and decompositions aforesaid are introduced into the components of the Einstein equation. Maintaining only the first-order terms, we obtain:
\begin{eqnarray}
&&\delta\lambda=-\frac{2re^{\lambda}}{d-3}S_{d-2}G_d\left(p_{d}+\rho_{d}\right)\zeta,\\
&&\frac{\partial\left(\delta\nu\right)}{\partial r}=\frac{2re^{\lambda}}{d-3}S_{d-2}G_d\left(\delta p_{d}\right.\nonumber\\
&&\left.-\left(p_{d}+\rho_{d}\right)\zeta\left(\frac{d\nu}{dr}+\frac{d-3}{r}\right)\right),\\
&&\delta\rho_{d}=-\frac{d\rho_{d}}{dr}\zeta-\frac{\left(p_{d}+\rho_{d}\right)e^{\nu/2}}{r^{d-2}}\frac{d}{dr}\left(\frac{r^{d-2}\zeta}{e^{\nu/2}}\right),\\
&&\delta p_{d}=-\frac{dp_{d}}{dr}\zeta-\frac{\Gamma_1 p_{d}e^{\nu/2}}{r^{d-2}}\frac{d}{dr}\left(\frac{r^{d-2}\zeta}{e^{\nu_0/2}}\right),
\end{eqnarray}
where $\Gamma_1=\left(\frac{p_{d}+\rho_{d}}{p_{d}}\right)\frac{dp_{d}}{d\rho_{d}}$ represents the adiabatic index and $\zeta$ depicts the ``Lagrangian displacement"  with respect to the world time $t$ defined by $v=\partial\zeta/\partial t$. To analyze the stability of hyperspherical objects against radial perturbations, the linearized form of the  energy-momentum tensor conservation and the aforementioned perturbed quantities with their temporal dependences of the form $e^{i\omega t}$ must be considered. Thus, we get:
\begin{eqnarray}\label{ROE}
&&\omega^2\left(p_{d}+\rho_{d}\right)e^{\lambda-\nu}\zeta-\left(\frac{2}{d-3}\right)S_{d-2}G_d\left(p_{d}+\rho_{d}\right)\zeta e^{\lambda}\,p_{d}\nonumber\\
&&+e^{-\lambda/2-\nu}\frac{d}{dr}\left(e^{\lambda/2+\nu}\frac{p_{d}\Gamma_1}{r^{d-2}}e^{\nu/2}\frac{d}{dr}\left(e^{-\nu/2}r^{d-2}\zeta\right)\right)\nonumber\\
&&-\frac{2(d-2)\zeta}{r}\frac{dp_{d}}{dr}+\frac{\left(p_{d}+\rho_{d}\right)\zeta}{4}\left(\frac{d\nu}{dr}\right)^2=0,
\end{eqnarray}
where $\omega$ is known as the eigenfrequency. Equation  \eqref{ROE} is known as the radial pulsation or Chandrasekhar pulsation equation. This equation is modified from its traditional form to analyze the influence of the extra dimensions in the radial stability. Moreover, Eq.~\eqref{ROE} could be placed in a more appropriate form to numerical integration. Following Ref.~\cite{gondek1997} (see also Ref.~\cite{lugones2010}), the radial oscillation equation is placed into the form of two first-order equations as shown below:
\begin{eqnarray}
\hspace{-0.3cm}&&\frac{d\xi}{dr}=\frac{\xi}{2}\frac{d\nu}{dr}-\frac{1}{r}\left((d-1)\xi+\frac{\Delta p_{d}}{p_{d}\Gamma_1}\right),\label{ROE_1}\\
\hspace{-0.3cm}&&\frac{d\Delta p_{d}}{dr}=\frac{\xi re^{\lambda}}{e^{\nu}}(p_{d}+\rho_{d})\omega^2+\frac{(p_{d}+\rho_{d})r\xi}{4}\left(\frac{d\nu}{dr}\right)^2\nonumber\\
\hspace{-0.3cm}&&-\left(S_{d-2}G_d\frac{re^{\lambda}(p_{d}+\rho_{d})}{d-3}+\frac{1}{2}\frac{d\nu}{dr}\right)\Delta p_{d}-2(d-2)\xi \frac{dp_{d}}{dr}\nonumber\\
\hspace{-0.3cm}&&-2S_{d-2}G_d(p_{d}+\rho_{d})e^{\lambda}r\xi\left(\frac{p_{d}}{d-3}\right),\label{ROE_2}
\end{eqnarray}
where $\xi=\zeta/r$ denotes the relative radial displacement and $\Delta p$ depicts the Lagrangian perturbation. For $d=4$, Eqs.~\eqref{ROE_1} and \eqref{ROE_2} are reduced to the two first-order equation form presented in Ref.~ \cite{gondek1997,lugones2010}.

To integrate the differential equations \eqref{ROE_1} and \eqref{ROE_2}, the boundary conditions must be defined. With the aim of determining regular solutions, in the center of the object ($r\to0$), the following is required:
\begin{equation}
\Delta p_{d}\,G_d=-(d-1)\left(\xi\Gamma_1\,p_{d}\,G_d\right)_{\rm center}.
\end{equation}
In $r\to0$, for normalized eigenfunctions, we have $\xi(r=0)=1$. On the other hand, the object's surface ($r=R$) is attained when $p_{d}\,G_d\to0$ and, consequently,
\begin{equation}
\left(\Delta p_{d}\,G_d\right)_{\rm surface}=0.\label{ext_cod_ro}
\end{equation}

\subsection{Equation of state}

For the strange quark matter contained in the object, we consider the pressure $p_{d}$ and energy density $\rho_{d}$ are connected through a generalization of the MIT bag model EOS in $d$-dimensions:
\begin{equation}\label{eos}
p_d=\frac{1}{d-1}\left(\rho_d-d\,{\cal B}_d\right),
\end{equation}
with ${\cal B}_d$ being the generalized bag constant. For a four-dimensional spacetime, $d=4$, this equality depicts a fluid made of up, down and strange quarks. It was proposed by Witten \cite{witten1984} that the strange quark matter could be the true ground state of strongly interacting matter. This conjecture is verified by Farhi and Jaffe \cite{farhi_jaffe1984} considering massless and noninteracting quarks. The four-dimensional MIT bag model equation of state was considered in some previous investigations on structure and stability against radial perturbations, for example, to investigate the radial pulsations in Refs.~ \cite{gondek1999,vath_chanmugam1992,benvenuto_horvath1991,arbanil_malheiro,arbanil_malheiro2016} and the stability of thin shell interfaces within compact stars \cite{pereira_coelho_rueda2014} (see also Refs.~\cite{pereira_rueda2015,pereira_flores2018}). 

Owing to the volume, the bag constant ${\cal B}_d$ and functions $\rho_d$ and $p_d$ are dimension dependent units. Thus, with the objective of having these units be spacetime dimension independent, we use those of the form ${\cal B}_d\,G_d$, $\rho_d\,G_d$ and  $p_d\,G_d$. The units of all of these variables are $[\rm MeV/fm^3]$.

With the aim of comparing our results with those ones obtained in the four-dimensional spacetime \cite{gondek1999,vath_chanmugam1992,benvenuto_horvath1991,arbanil_malheiro,arbanil_malheiro2016}, we use Eq.~\eqref{eos} by considering $d\,{\cal B}_d\,G_d= 240\,[\rm MeV/fm^3]$.

\section{Influence of the dimension in the equilibrium and stability of compact objects}\label{results}

\subsection{Numerical method}

To investigate the extra dimensions influence on the static equilibrium configurations and radial stability, the stellar structure and radial oscillation equations are, respectively, resolved. These equations are integrated from the center toward the surface of the object. 

\subsubsection{Numerical method for the stellar structure equations}

For the hydrostatic equilibrium, the stellar structure equations, Eqs.~\eqref{df/f}, \eqref{dm/dr} and \eqref{tov}, equation of state \eqref{eos} and boundary conditions \eqref{int_cond} and \eqref{pressure_r} are numerically solved. This is realized through the fourth-order Runge-Kutta method complemented with the shooting method.

We solve Eqs. \eqref{dm/dr}, \eqref{tov}, and \eqref{eos} using the fourth-order Runge-Kutta method for some $\rho_{cd}\,G_d$ and $d$. Once we obtain $p_d\,G_d$, $\rho_d\,G_d$, $m\,G_d$ and $\lambda$ for a given $\rho_{cd}\,G_d$ and $d$, Eq.~\eqref{df/f} is solved the by mean of the shooting method. To solve numerically solve this equation, first, we consider a trial value for $\nu_c$. If after the integration the boundary condition \eqref{metric_functions} is not fulfilled, we repeat this process until satisfy it is satisfied.

\subsubsection{Numerical method for the radial stability equations}

Equations of radial pulsations are solved using the shooting method. This method starts by considering the coefficients of the radial oscillation equations, which are obtained solving the stellar structure equations for a specific $\rho_{cd}\,G_d$ and $d$ and a proof value for $\omega^2$. If at the end of the numerical integration, the boundary condition \eqref{ext_cod_ro} is not satisfied, $\omega^2$ is rectified untilit is attained in the next integration.

\subsection{Influence of the dimension in the pressure and energy density}

\begin{figure}[ht]
\begin{center}
\includegraphics[width=0.97\linewidth]{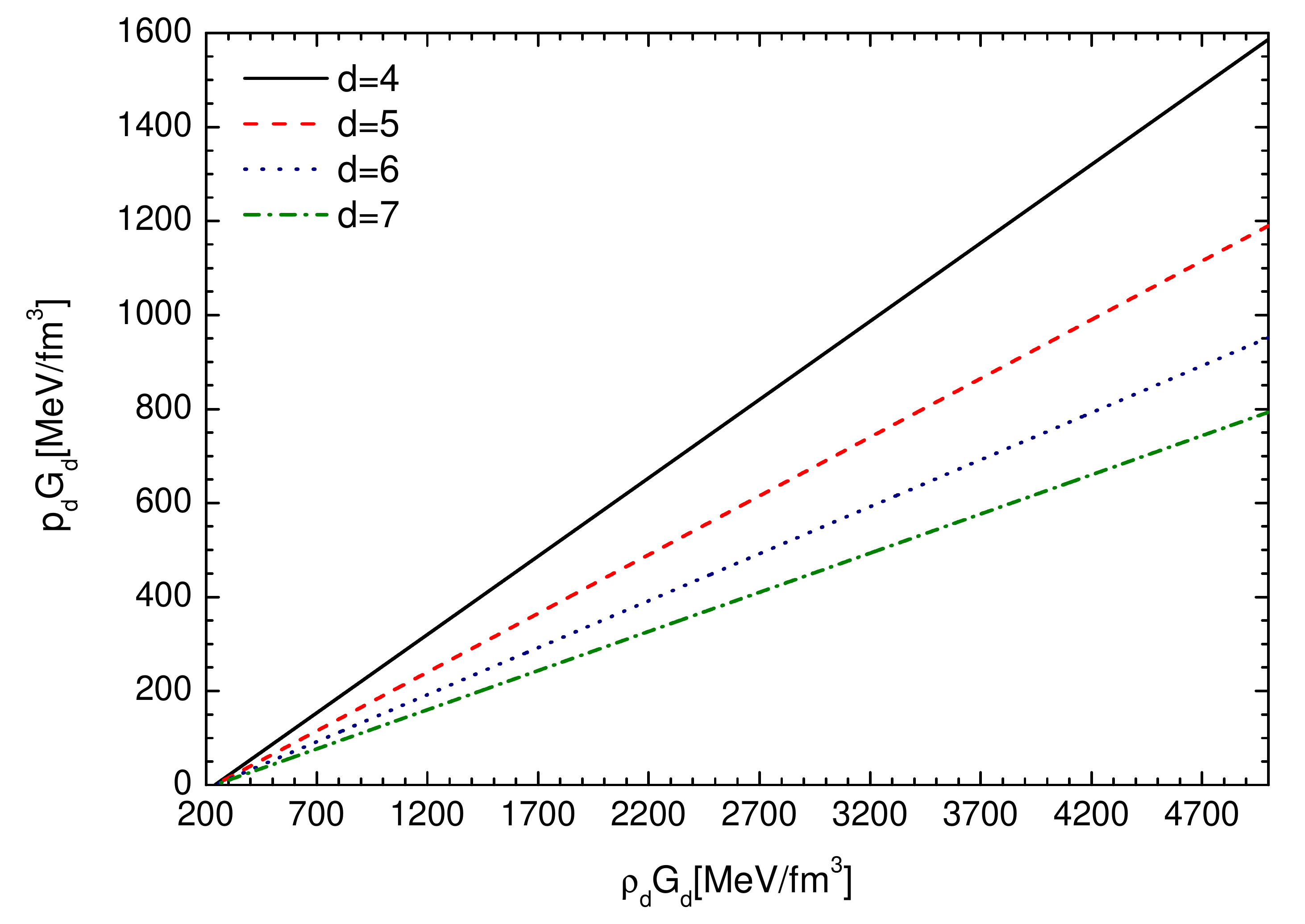}
\caption{Pressure versus energy density for a few different spacetime dimensions.}
\label{p_rho}
\end{center}
\end{figure}

In order to check the EOS behavior, in Fig.~\ref{p_rho}, the pressure $p_d\,G_d$ against energy density $\rho_d\,G_d$ is plotted for some spacetime dimensions. The energy density considered runs from $240\,[\rm MeV/fm^3]$ ($d\,{\cal B}_d\,G_d$) to $5000\,[\rm MeV/fm^3]$. It is observed that a linear EOS in four dimensions remains linear in $d>4$. From graphic, also note that the fluid pressure decays with the increment of $d$.


\subsection{Static equilibrium configurations in a $d$-dimensional spacetime}

\begin{figure}[ht]
\begin{center}
\includegraphics[width=0.97\linewidth]{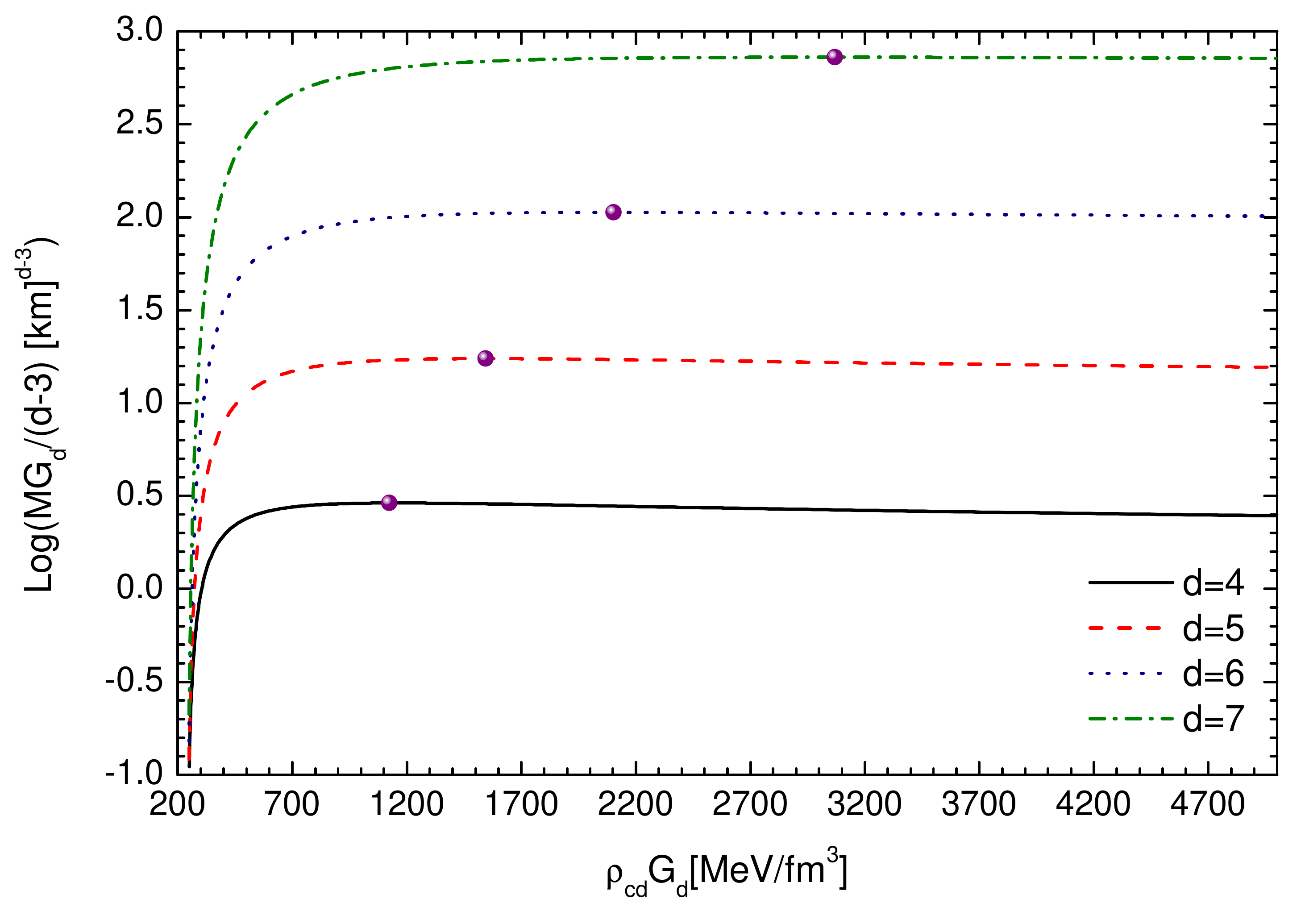}
\caption{Mass of the object against the central energy density for different spacetime dimensions. The full circles indicate the places where maximum masses are found.}
\label{M_rhoc}
\end{center}
\end{figure}

The total mass of the object $M\,G_d/(d-3)$ versus the central energy density $\rho_{cd}\,G_d$ is plotted in Fig.~\ref{M_rhoc} for different spacetime dimensions. The central energy density used goes from $250$ to $5000\,[\rm MeV/fm^3]$. The complete circles established on the curves denotes the places where maximum masses are found. It is important to say that the curve built for a four-dimensional spacetime ($d=4$), is similar to the one found in Refs.~ \cite{arbanil_malheiro,arbanil_malheiro2016}. In all spacetime dimensions considered, the mass increases with the central energy density until the maximum mass point is attained; henceforward, the mass decreases with the increment of $\rho_{cd}\,G_d$. In addition, note that the mass is nearly constant for central energy densities larger than $\sim700\,[\rm MeV/fm^3]$. 

Figure \ref{M_rhoc} also shows the influence of the spacetime dimension $d$ in the mass of compact objects. In order to compare the masses of the objects found in higher dimensions with those obtained in a four-dimensional spacetime, we consider the length of the extra dimension around $\ell\sim10^{-18}[\rm km]$; see Ref.~\cite{antoniadins2010}. Thus, from Table I, we can distinguish that the maximum mass $(M\,G_d)_{\rm Max}$ found in each spacetime dimension, $d>4$, with respect the one found in $d=4$, changes roughly with the factor:
\begin{equation}
\frac{\left[M_d\right]_{\rm Max}}{\left[M_4\right]_{\rm Max}}\sim \left[10^{19}\right]^{d-4}.
\end{equation}
The last relation indicates that the maximum masses of objects in dimensions $d>4$ are much larger than the ones determined in $d=4$. Whether the extra dimension length is larger than the Planck length, for $d>4$, compact objects with smaller masses can be found. In addition, it is important to say that the compact objects with masses $M\,G_d\leq (M\,G_d)_{\rm Max}$ are stables against small radial perturbations (see e.g. Fig.~\ref{omega_m}). Contrary to what is obtained in Ref.~\cite{chavanis2007}, because the energy density is non-null at the surface the object, we found stable compact stars in $d>4$ .

\begin{figure}[ht]
\begin{center}
\includegraphics[width=0.97\linewidth]{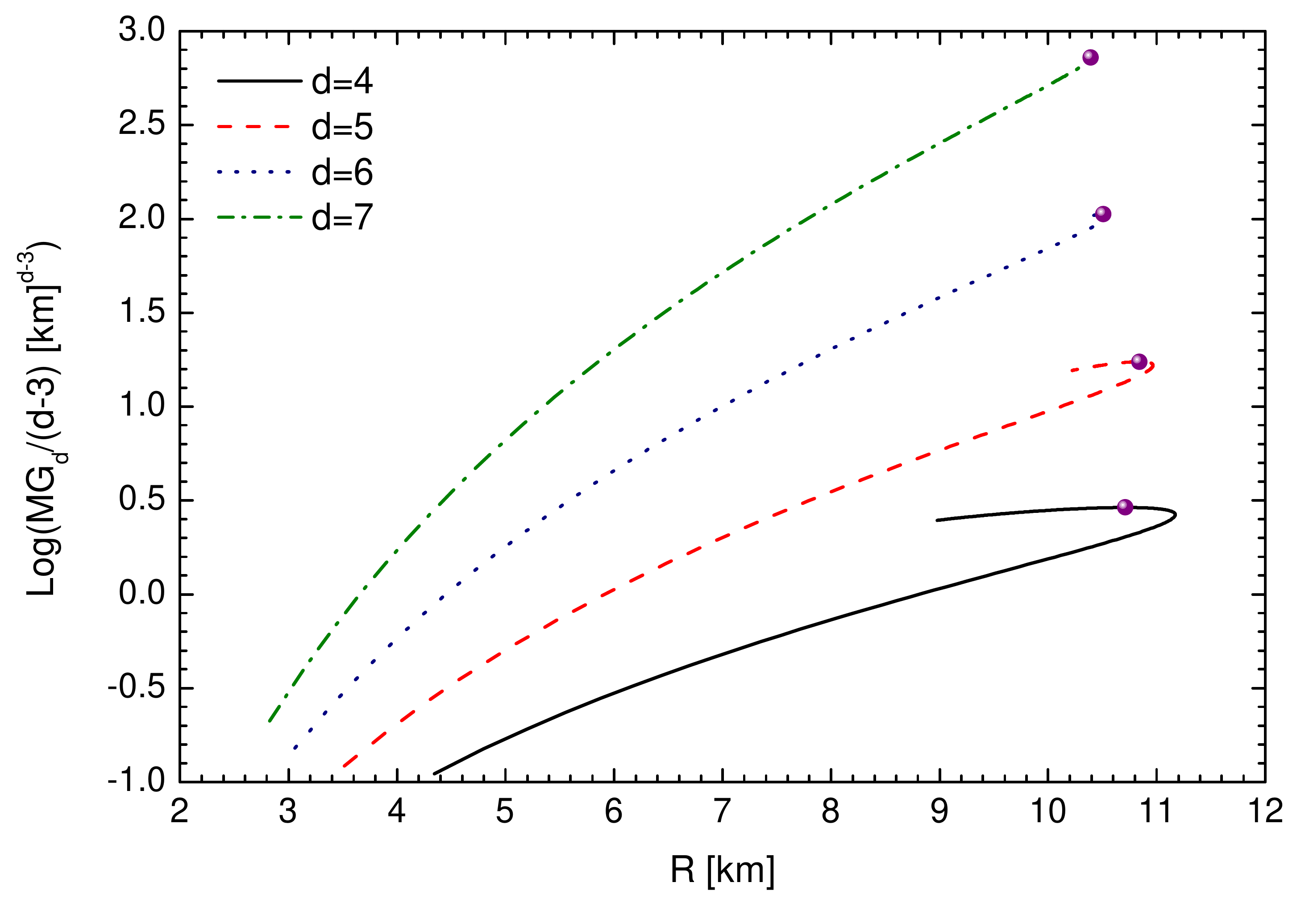}
\caption{Total mass of the object against the total radius for different spacetime dimensions. The complete circles mark the points where the maximum masses are found.}
\label{r_m}
\end{center}
\end{figure}

The total mass as a function of the total radius is presented in Fig.~\ref{r_m} for different spacetime dimensions. As in Fig.~\ref{M_rhoc}, the complete circles over the curves point out the places where maximum masses are found. In the figure, we see how both the mass and the radius change with the spacetime dimension. In four dimensions, as it is characteristic in strange stars, in the interval $2.22\lesssim M\lesssim 2.95\,[\rm km]$, the masses and radii follow roughly the relation $M(R)\propto R^3$. In addition, it is important to say that the curve bends counterclockwise for a total mass of around $2.95\,[\rm km]$ (for a review, see for instance, Ref.~ \cite{lugones_arbanil2017}). 

\begin{figure}[ht]
\begin{center}
\includegraphics[width=0.97\linewidth]{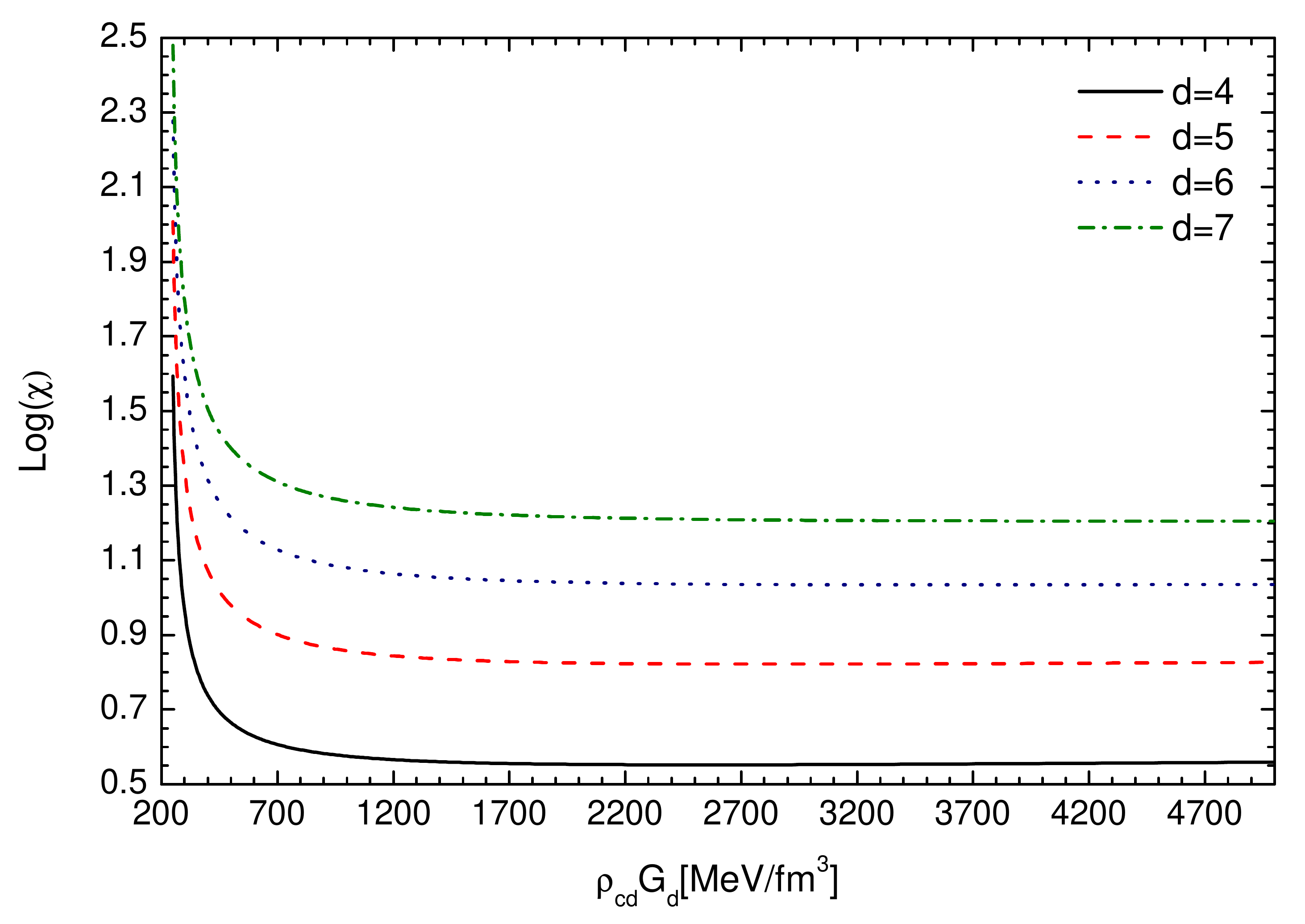}
\caption{Ratio $(d-3)R^{d-3}/MG_d=\chi$ as a function of the energy density for different spacetime dimensions.}
\label{RM_rhoc}
\end{center}
\end{figure}

The ratio ${(d-3)R^{d-3}}/{M\,G_d}=\chi$ as a function of the central energy density $\rho_{cd}\,G_d$ is presented in Fig.~\ref{RM_rhoc} for some spacetime dimensions $d$. In all curves, note that $\chi$ decreases monotonically with the central energy density until reaching $\rho_{cd}\,G_d\sim1200\,[\rm MeV/fm^{3}]$, hereafter, the factor $\chi$ is nearly constant. This implies that, for central energy densities larger than $1200\,[\rm MeV/fm^3]$, $\chi$ is independent of $\rho_{cd}\,G_d$.

The dependence of ${(d-3)R^{d-3}}/{M\,G_d}$ with the spacetime dimension $d$ is also shown in Fig.~\ref{RM_rhoc}. In all cases, $\chi$ is far from attaining the Buchdahl limit \cite{buchdahl} for a $d$-dimensional spacetime; i.e., the factor $(d-1)^2/2(d-2)$ is far from being reached \cite{dadhichPRD,poncedeleon2000}.  

\begin{figure}[ht]
\begin{center}
\includegraphics[width=0.97\linewidth]{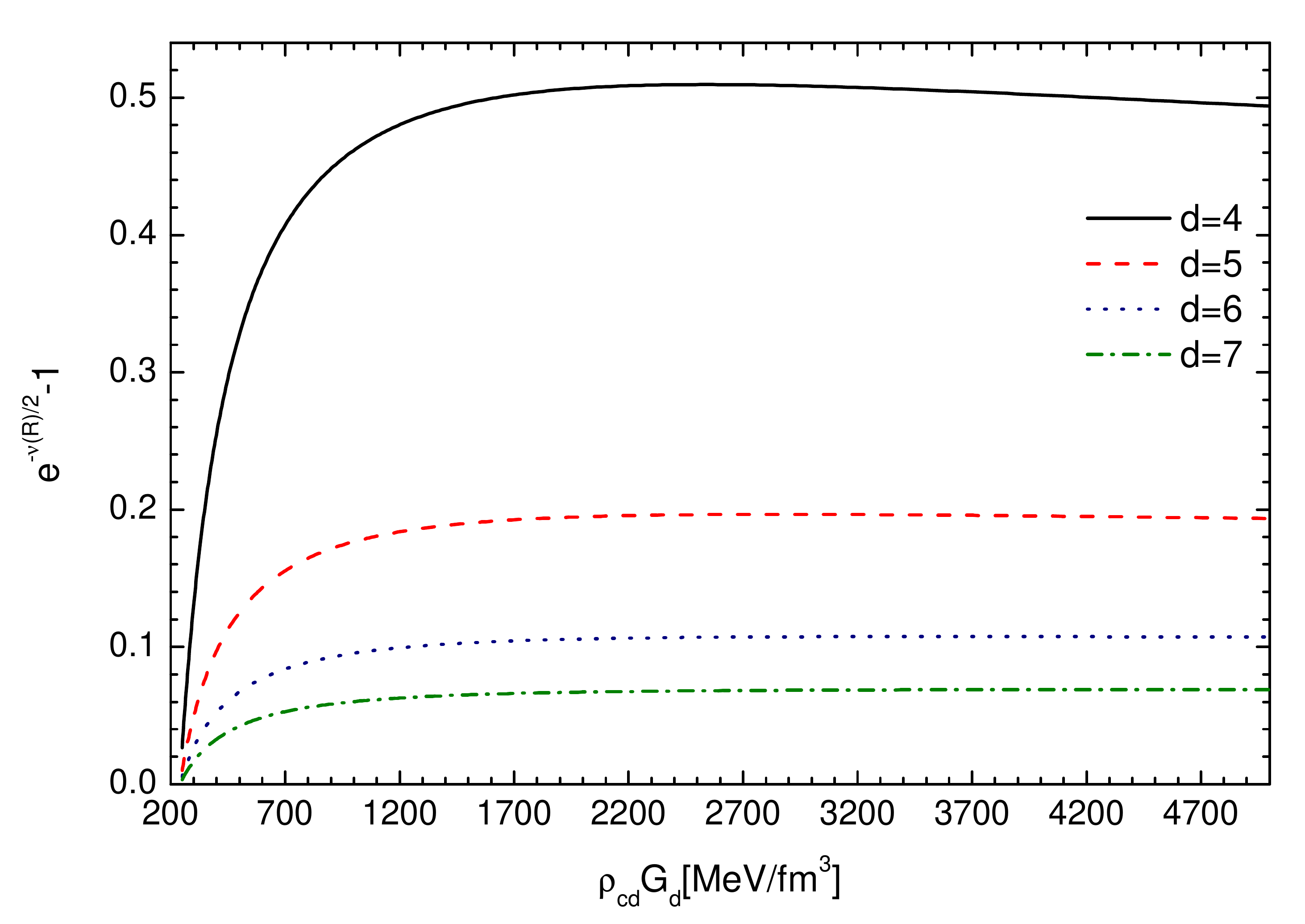}
\caption{The redshift function at the object's surface, $e^{-\nu(R)/2}-1$, as a function of the central energy density, for different spacetime dimensions.}
\label{RS_rhoc}
\end{center}
\end{figure}

An important factor to be investigated is the redshift at the object's surface. It shows how the light emitted by an object is deflected to the red end of the spectrum. Thus, the redshift at the surface of the star versus the central energy density is investigated in Fig.~\ref{RS_rhoc} for different spacetime dimensions. For $d\leq6$, the redshift grows with $\rho_{cd}\,G_d$ up to a point of inflection. After this peak, the redshift begins to fall with the increase of the central energy density. In turn, for $d>6$, the redshift grows monotonically with $\rho_{cd}\,G_d$. In the last case, a maximum redshift point is not found; moreover, when $\rho_{cd}\,G_d\sim1200\,[\rm MeV/fm^3]$, the redshift remains nearly constant.

In addition, it is clearly noted in Fig.~\ref{RS_rhoc} that the redshift changes with the dimension. For  larger dimensions, compact objects with lower redshifts are found. 

\begin{table}
\begin{ruledtabular}
\begin{tabular}{ccccccc}
$d$ & $MG_d/(d-3)$ & $\rho_{cd}\,G_d$& $R$ & $\chi$ \\\hline
$4$ & $2.8997$     & $1123.2$ & $10.710$ & $3.6935$ \\
$5$ & $17.321$     & $1554.4$ & $10.842$ & $6.7865$ \\
$6$ & $106.06$     & $2102.4$ & $10.510$ & $10.946$ \\
$7$ & $723.54$     & $2570.7$ & $10.391$ & $16.113$ 
\end{tabular}
\caption{Spacetime dimensions and the maximum values of $MG_d/(d-3)$ with their respective central energy densities $\rho_{cd}\,G_d$, radii $R$ and ratios $\chi\equiv(d-3)R^{d-3}/M\,G_d$. The units of the maximum masses, the central energy densities and radii are respectively $[\rm km]^{d-3}$, $[\rm MeV/fm^3]$ and $[\rm km]$.}
\end{ruledtabular}\label{table}
\end{table}

In Table I is presented the spacetime dimensions used and the maximum value of $MG_d/(d-3)$ with their respective central energy densities $\rho_{cd}\,G_d$, total radii $R$, and compressibility factors $\chi$.

\subsection{Radial stability in a $d$-dimensional spacetime}

\begin{figure}[ht]
\begin{center}
\includegraphics[width=0.97\linewidth]{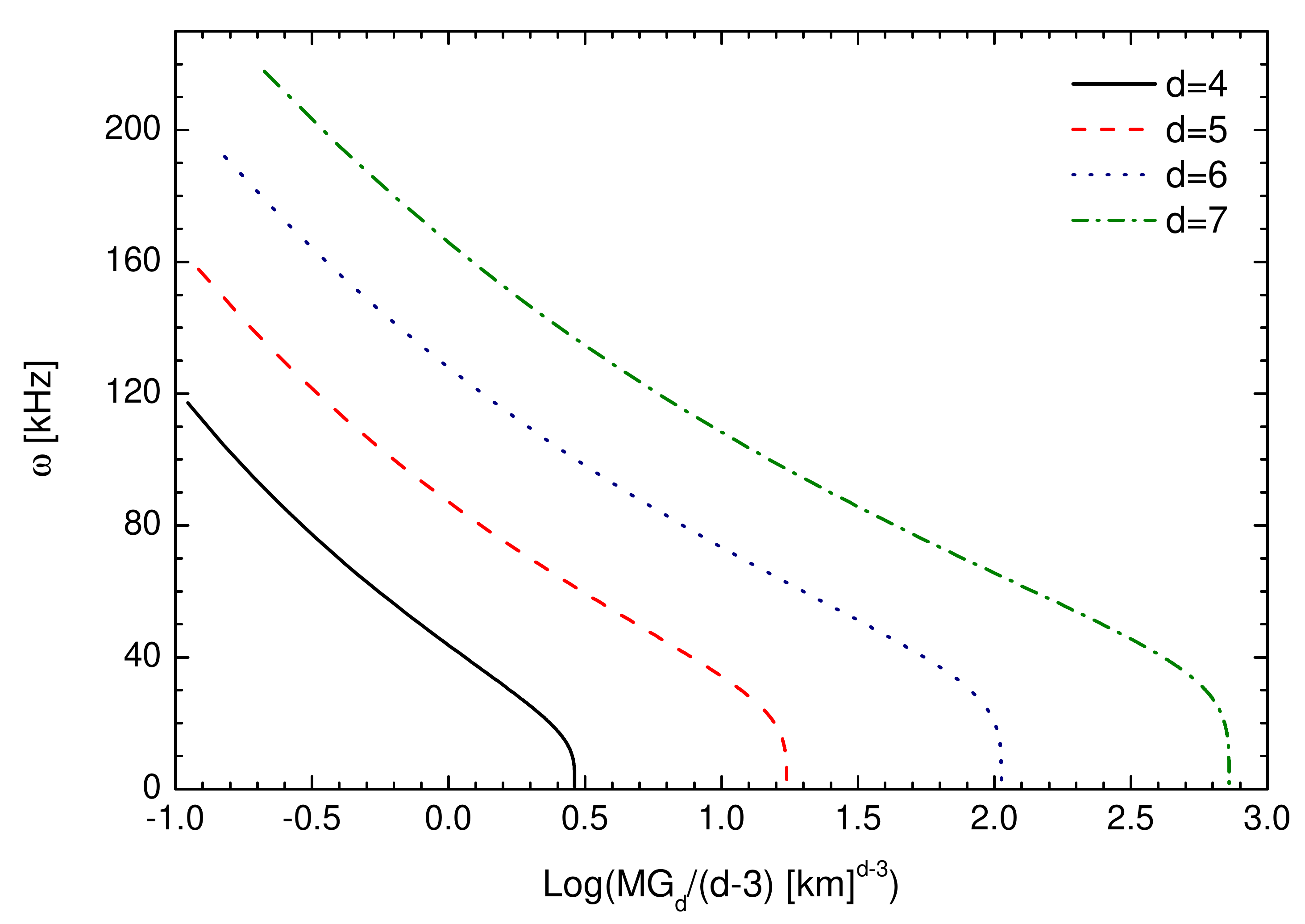}
\caption{Fundamental mode eigenfrequency versus the mass of the object for different dimensions.}
\label{omega_m}
\end{center}
\end{figure}

The behavior of the fundamental mode's eigenfrequency with the total mass is presented in Fig.~\ref{omega_m} for different dimensions. In the figure, we consider static equilibrium configurations with $\omega\geq0$. It is important to say that the behavior of the curve found in $d=4$ is very similar to the one derived in the study of radial stability of strange stars in Refs.~\cite{gondek1999,arbanil_malheiro,arbanil_malheiro2016}. In all curves, a monotonic decline of $\omega$ with the increment of the mass is observed until attain a zero eigenfrequency in the maximum total mass value. Thus, independent of the spacetime dimension, the maximum mass point $dM/d\rho_{cd}=0$ marks the onset of instability; see Fig.~\ref{M_rhoc}. In other words, in a sequence of compact objects in the same spacetime dimension, the conditions $dM/d\rho_{cd}>0$ and $dM/\rho_{cd}< 0$ are necessary and sufficient to identify regions made of stable and unstable stars against radial perturbations, respectively. 

In addition, the radial stability of compact objects in a higher-dimensional spacetime is analyzed in a similar way as the turning-point method for axisymmetric stability of rotating relativistic stars Ref.~\cite{friedman1988,sorkin1982} (for a detailed discussion about this point, see \cite{takami2011}). Instead of fixing the angular momentum in a sequence of rotating compact objects to calculate the turning point, in this work, the spacetime dimension is fixed in a sequence of spherically symmetric static objects.

On the other hand, the influence of the dimension is viewed in Fig.~\ref{omega_m}. For some interval of mass, the increment of the dimension helps to grow the radial stability.

\begin{figure}[ht]
\begin{center}
\includegraphics[width=0.97\linewidth]{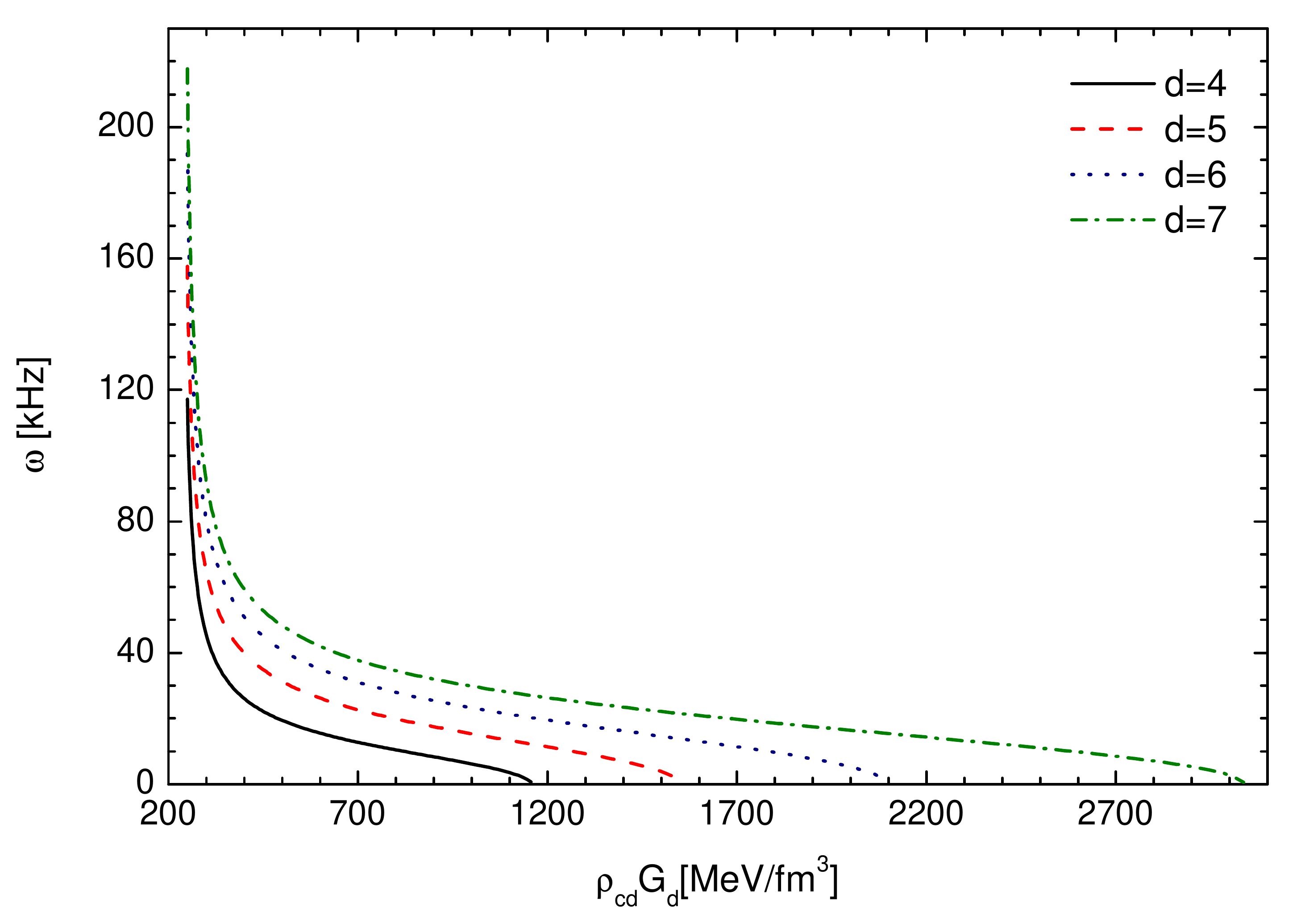}
\caption{Fundamental mode eigenfrequency against the central energy density for different dimensions.}
\label{omega_rhoc}
\end{center}
\end{figure}

The eigenfrequency of the fundamental mode of oscillation against central energy density is plotted in Fig.~\ref{omega_rhoc} for different spacetime dimensions. As noted in the figure, we only consider $\rho_{cd}\,G_d$ with $\omega\geq0$. In all cases, $\omega$ diminishes monotonically with central energy density until it reaches the zero eigenfrequency; this indicates that the radial stability of a compact object decreases with the increment of the central energy density.

The effect of the dimension on the radial stability is also noted in Fig.~\ref{omega_rhoc}. For a fixed central energy density, the fundamental eigenfrequency of oscillation $\omega$ grows with the dimension $d$; i.e., an object is more stable in higher dimensions.

\begin{figure}[ht]
\begin{center}
\includegraphics[width=0.97\linewidth]{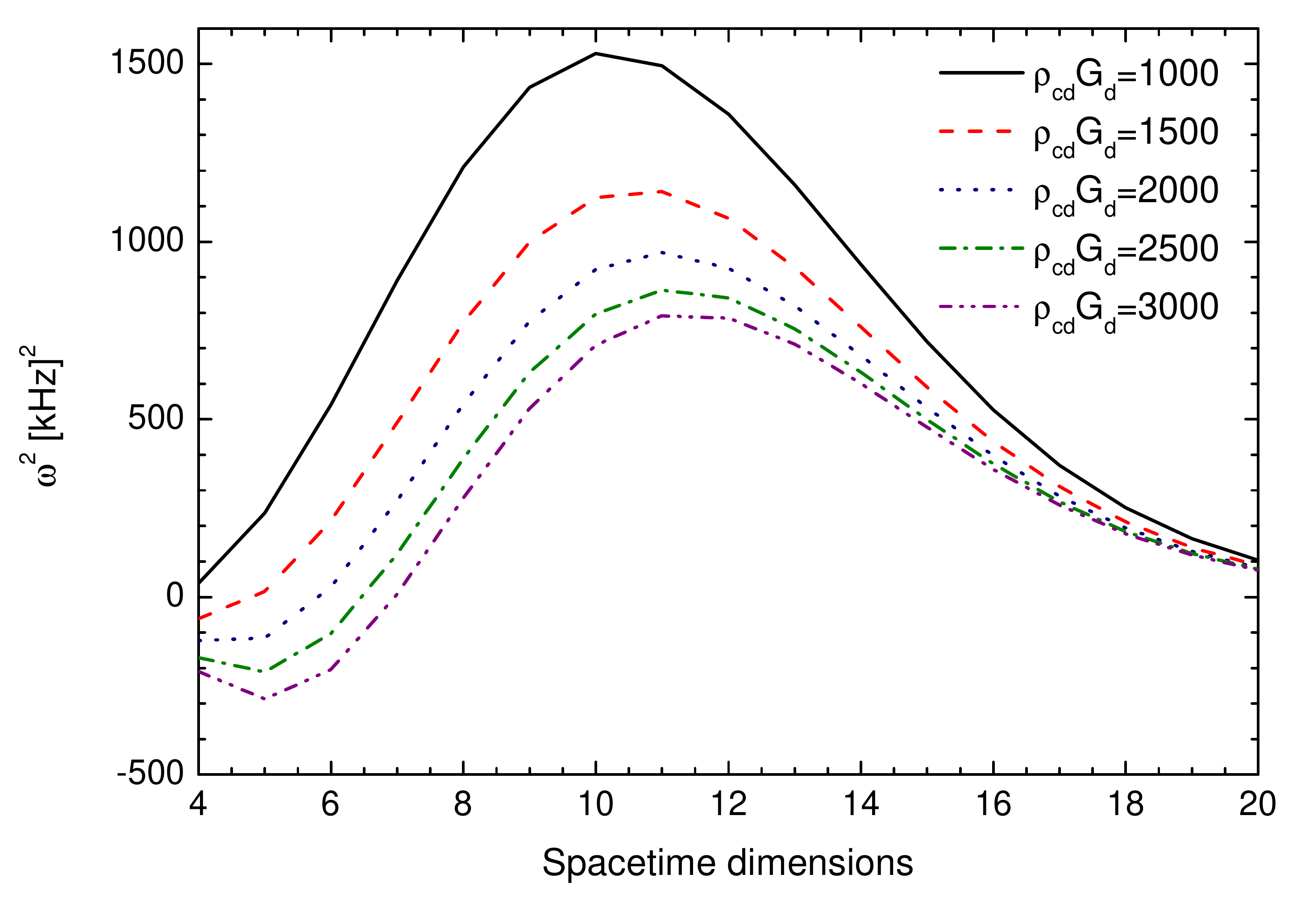}
\caption{Change of the fundamental mode eigenfrequency of oscillation squared with the spacetime dimensions for five different central energy densities. The central energy density units are $[\rm MeV/fm^3]$.}
\label{w2_d}
\end{center}
\end{figure}

In Fig.~\ref{w2_d} is schemed the  squared eigenfrequency of the fundamental mode against the spacetime dimension  for five different central energy densities. From graphic, the change of the radial stability with spacetime dimensions  is observed. For objects with $\rho_{cd}\,G_d$ lower than $\sim2000\,[\rm MeV/fm^3]$, the radial stability increases with the dimension until attain a turning point in $d=10$; hereafter, $\omega^2$ starts decreasing with $d$. In turn, for $\rho_{cd}\,G_d\gtrsim2000\,[\rm MeV/fm^3]$, the stability of the object decreases when the dimension goes from $4$ to $5$. From $d\geq5$, the eigenfrequency of oscillation increments with the dimension until $d=11$; henceforth, the eigenfrequency of the fundamental mode decays with the spacetime dimensions.

\subsection{About the $d$-dimensional equilibrium and stability for other normalizations}

The stellar equilibrium equations derived in this work and those ones found by Harko and Mak \cite{harko_mak2000} and Ponce de Leon and Cruz \cite{poncedeleon2000} are obtained by using different normalizations. The form of the stellar structure equations in each work depend on how both the Einstein field equation and the $d$-dimensional mass are defined.

By using scales of Table II within Eqs.~\eqref{dm/dr} and \eqref{tov}, equation of structures presented in \cite{harko_mak2000} and \cite{poncedeleon2000} are recovered. This point outs that the inputs values used and the total masses defined in this article must change. This change can be realized using the relations presented in Table II. 

\begin{table}[htb]
\begin{tabular}{l||ll|ll|l}
                                & \multicolumn{2}{c|}{Scales}  & \multicolumn{2}{c|}{Inputs} & \multicolumn{1}{c}{Masses}\\\hline\hline
ACLMM & $G_d$ & $m\,G_d$            & $\rho_{cd}\,G_d$                & ${\cal B}_d\,G_d$                        & $M\,G_d$\\
HM            & ${\tilde G}_d\,a_1$ & ${\tilde m}\,{\tilde G}_d\,b_2$ & ${\tilde \rho}_{cd}\,{\tilde G}_d\,a_1$  & ${\tilde {\cal B}}_d\,{\tilde G}_d\,a_1$    & ${\tilde M}\,{\tilde G}_d\,b_2$\\
PC            & ${\bar G}_d\,a_1$   & ${\bar m}\,{\bar G}_d\,b_1$     & ${\bar \rho}_{cd}\,{\bar G}_d\,a_1$      & ${\bar {\cal B}}_d\,{\bar G}_d\,a_1$    & ${\bar M}\,{\bar G}_d\,b_1$
\end{tabular}
\caption{Normalizations used to transform both the stellar structure equations and results presented by Arba\~nil, Carvalho, Lobato, Marinho and Malheiro (ACLMM) into the formalism used by Harko and Mak (HM) \cite{harko_mak2000} and Ponce de Leon and Cruz (PC) \cite{poncedeleon2000}. The tildes and bars over some constants and variables are used to distinguish between the ones taken into account by HM and PC, respectively. The constants $a_1=8\pi(d-3)/S_{d-2}(d-2)$, $b_1=(d-3)$, $b_2=(d-3)/(d-2)2^{d-5}$.}\label{table2}
\end{table}

Albeit the normalizations used in Refs.~\cite{harko_mak2000} and \cite{poncedeleon2000} are different,  for a central energy density $\rho_c^*$ and a spacetime dimension $d$, the radius and mass are in both cases the same. However, in this work, for a central energy density $\rho_c^*$ and a spacetime dimension $d$, the radius and mass are smaller than those ones found in Refs.~ \cite{poncedeleon2000,harko_mak2000}. In addition, independent of the normalization used, the zero eigenfrequency of oscillations and the maximum masses values are found in the same central energy densities; in this way, the maximum mass marks the beginning of instability.

\subsection{Perturbation about equilibrium in Newtonian limit}

In the Newtonian limit, the EOS \eqref{eos} is well represented by the EOS $\rho_d=d{\cal B}_d$. Note that this EOS describes a constant energy density along the whole compact object. In these homogeneous objects in equilibrium, the fluid pressure is given by 
\begin{equation}\label{p_newtonian}
p_d=\frac{S_{d-2}G_d(d{\cal B}_d)^2}{2(d-1)}\left(R^2-r^2\right).
\end{equation}
To analyze the radial stability of homogeneous objects in the Newtonian limit, Eq.~\eqref{ROE} takes the form
\begin{equation}\label{ROE22}
\omega^2\rho_d\zeta+\frac{d}{dr}\left[\frac{p_d\Gamma_1}{r^{d-2}}\frac{d}{dr}\left[r^{d-2}\zeta\right]\right]-\frac{2(d-2)\zeta}{r}\frac{dp}{dr}=0.
\end{equation}
From this last equation, regular solutions are obtained since
\begin{eqnarray}
&&\zeta=0\;\;\;\;\;\;\;{\rm at}\;\;\;\;\;r=0,\label{bc_nc_1}\\
&&\zeta={\rm finite\;\;at}\;\;\;\;\;r=R.\label{bc_nc_2}
\end{eqnarray}

Substituting Eq.~\eqref{p_newtonian} into Eq.~\eqref{ROE22}, we obtain
\begin{equation}\label{eq_equilibrium_newtonian}
(1-x^2)\zeta''+\left[\frac{d-2}{x}-dx\right]\zeta'+\left[A_0-\frac{d-2}{x^2}\right]\zeta.
\end{equation}
In this last equation, the dimensionless function $x=\frac{r}{R}$, the prime over the variables represents $(')=\frac{d}{dx}$ and
\begin{equation}
A_0=\frac{2(d-1)\omega^2}{S_{d-2}G_d(d{\cal B}_d)\Gamma_1}-(d-2)+\frac{4(d-2)}{\Gamma_1}.
\end{equation}
In order to solve Eq.~\eqref{eq_equilibrium_newtonian} and to satisfy boundary condition \eqref{bc_nc_1}, following Ref.~\cite{shapiro2008}, we consider $\zeta$ of the form
\begin{equation}\label{serie_new}
\zeta=\sum_{n=0}^{\infty}a_{n}x^{n+1}.
\end{equation}
Substituying Eq.~\eqref{serie_new} in Eq.~ \eqref{eq_equilibrium_newtonian}, after some algebra, we find that $a_1=a_3=a_5=0$ and 
\begin{equation}
\frac{a_{n+2}}{a_n}=\frac{n^2+(d+1)n+d-A_0}{n^2+(d+3)n+2d+2},\;\;\;n=0,2,4,\hdots.
\end{equation}
From this last equation, the series diverges and, consequently, $\zeta$ does not satisfy condition \eqref{bc_nc_2}. Thus, with the aim of obtaining $\zeta={\rm finite}$ at $r=R$, it is required that
\begin{equation}
A_0=n^2+(d+1)n+d,\;\;\;\;\;n=0,2,4,\hdots,
\end{equation}
and, as a consequence,
\begin{equation}\label{w2_newtonian}
\begin{array}{r}
\omega^2=\frac{S_{d-2}G_d\,d\,{\cal B}_d}{2(d-1)}\left[\Gamma_1\left(n^2+(d+1)n+2(d-1)\right)\right.\\\left.-4(d-2)\right],\;\;\;\;\;n=0,2,4,\hdots.
\end{array}
\end{equation}
From this last equality, the compact objects are stable if
\begin{equation}
\Gamma_1\geq\frac{2(d-2)}{(d-1)}.
\end{equation}
In addition, from Eq.~\eqref{w2_newtonian}, the neutral equilibrium mode $\omega^2=0$ is obtained with $\Gamma_1=2(d-2)/(d-1)$ and $n=0$.

It is important to say that, unlike what is obtained in white dwarf in $d$ dimensions, we found stable compact objects against radial perturbations. This is possible because, in the present case, their energy densities are nonzero at their surfaces.

\section{Conclusions}\label{conclusion}

The influence of the spacetime dimension in the equilibrium and radial stability of compact objects is investigated in this work. To this aim, the stellar structure equations and the Chandrasekhar radial pulsation equation are modified to include the extra dimensions effects. Moreover, it a linear relation between the pressure and energy density of the fluid is assumed. We consider that the equilibrium configurations investigated are matched smoothly with the exterior Schwarzschild-Tangherlini metric. Both the equilibrium and the stability are analyzed for different central energy densities $\rho_{cd}\,G_d$ and spacetime dimensions $d$.

Regarding the static equilibrium configurations, we note that some physical properties change with the spacetime dimension; namely, the fluid pressure, total mass, radius and redshift on the surface of the object depending on $d$. Considering that the length of the  extra dimension is around $\ell\sim10^{-18}[\rm km]$ (see Ref.~\cite{antoniadins2010}), the compact object maximum mass in each spacetime dimension, $d>4$, is $\sim[10^{19}]^{d-4}$ times the one found in $d=4$.

By means of the radial perturbation method, we find that, for a central energy density interval and total mass range, the increment of the spacetime dimension assist in enhancing the stability of compact stars. In addition, the maximum mass point and the zero fundamental eigenfrequency of oscillation are derived using the same central energy density $\rho_{cd}\,G_d$. From this, we infer that the conditions $dM/d\rho_{cd}> 0$ and $dM/d\rho_{cd}< 0$ are necessary and sufficient to distinguish, respectively, stable and unstable equilibrium configurations.

It is worth highlighting that the very massive and stable compact objects analyzed in this study might be hidden within the category of supermassive black holes. Certainly, in future, the results on the shadow of the M$87$ black hole \cite{event_horizon_collaboration/2019} could aid in discriminating these compact objects from black holes. In addition, if a very massive compact object is detected, it might be considered evidence of the existence of extra dimensions. 

We also find that some physical properties depend on the normalization considered. For a fixed central energy density $\rho_c^*$ and spacetime dimension $d$, we find a lower mass and total radius than those  derived in Refs.~\cite{poncedeleon2000,harko_mak2000}. In addition, in all spacetime dimensions used, independently of the normalization considered, we prove that the maximum mass point and the zero eigenfrequency of oscillation are found in the same central energy density. 

Finally, we investigate the radial stability of compact objects Newtonian regime. In this limit, the generalized MIT bag model EOS takes the form established for a homogeneous object. In contrast to what it is observed in white dwarfs \cite{chavanis2007}, we find stable compact objects against small radial perturbation. From this, we show that stable solutions for extradimensional spacetimes with the chosen EOS are obtained. We known that in a four-dimensional spacetime the effects of general relativity on the structure of white dwarfs are little dominant in their mass \cite{carvalho2018}. If we assume  that these effects are still small in more dimensions, white dwarfs within the  framework of general relativity in dimensions greater than $4$ are also unstable. Thus, since stars with the white dwarf EOS are unstable in extra dimensions, their astronomical observation exclude the possibility of an extended higher dimension. To reconcile this fact with the stable solution, we find,for extradimensional quark stars, that we need to consider that our extra dimensions need to be compactified. In fact, the quark confinement understood as a manifestation of quarks propagating in extra dimensions compactified is considered in the literature to explain color confining and the hadronic spectroscopy in the AdS-QCD conjecture \cite{bechhoefer_chabrier1993}. In our $d$-dimension strange stars the deconfinement density energy needed to liberate quarks from the confinement is in fact present, since the lowest energy density of the star surface  is $\rho_d=d{\cal B}_d$. This is exactly the deconfinement energy density in $d$ dimensions.

\begin{acknowledgments}
\noindent We would like to thank Funda\c{c}\~ao de Amparo \`a Pesquisa do S\~ao Paulo-FAPESP, Grant No. $2013/26258-4$. JDVA thanks Vilson T. Zanchin for the enriching discussions about the influence of higher dimensions in the configuration of some compact objects. GAC and RVL thank Coordena\c c\~ao de Aperfei\c coamento de Pessoal de N\'ivel Superior-CAPES for the Grants No. CAPES/PDSE/$88881.188302/2018-01$ and CAPES/PDSE/$88881.134089/2016-01$, respectively. RVL also thanks Conselho Nacional de Desenvolvimeno Cient\'ifico e Tecnol\'ogico-CNPq, Grant No. $141157/2015-1$.
\end{acknowledgments}

\end{document}